\documentclass[fleqn,twoside]{article}
\usepackage{espcrc2}

\usepackage{graphicx}
\usepackage[figuresright]{rotating}

\newcommand{\AmS}{{\protect\the\textfont2
  A\kern-.1667em\lower.5ex\hbox{M}\kern-.125emS}}

\title{The improved 10th order QED expresion for $a_{\mu}$:
new results and related estimates}

\author{A.~L.~Kataev\address[MCSD]{Institute for Nuclear Research of 
the Russian Academy of Sciences,\\
117312 Moscow, Russia}
\thanks{Supported by RFBR Grants N 03-02-17177, 05-01-00992}}
\begin{document}

\begin{abstract}
New  estimates of the 10th order QED  corrections 
to the  muon anomalous magnetic moment are presented. The  estimates    
include the information on    
definite improved 10th order QED contributions 
to $a_{\mu}$, calculated   by Kinoshita and Nio. 
The final estimates  are in good  agreement with the ones, 
given recently by Kinoshita. 
\vspace{1pc}
\end{abstract}
\maketitle

\section{INTRODUCTION}
In the last  years both theoretical and experimental 
results for the anomalous magnetic moment $a_{\mu}$ attracted special 
interest 
(for the most recent review see Ref. \cite{Passera:2004bj}).

Careful analysis  of the values of   
different theoretical  corrections  to 
$a_{\mu}$ stimulated the new fresh glance on the pure QED 
expression  for   this classical quantity. The  work was started 
after definite bugs in the previous 
calculations of  eighth-order light-by-light-type 
diagrams \cite{Kinoshita:1990wp}  were detected and corrected
\cite{Kinoshita:2002ns}. The evaluations of all 
mass-dependent $\alpha^4$ QED contributions to $a_{\mu}$ were 
completed in Ref. \cite{Kinoshita:2004wi} and their numerical 
values have been greatly improved with respect to 
previous results of Ref. \cite{Kinoshita:1990wp}. 

Moreover, the crude estimate of the $\alpha^5$ QED  
correction to $a_{\mu}$, which is  based on the calculations 
of the dominant contributions to the  sets of 
10th order light-by-light-type diagrams (see  
Ref. \cite{Kinoshita:1990wp}
and Ref. \cite{Milstein}) 
and the renormalization group inspired  studies  
of Refs. \cite{Kinoshita:1990wp,Karshenboim:1993rt} 
was also improved \cite{Kinoshita:2005ti}.
In view of this it is worthwhile to reconsider the 10th order 
scheme-invariant estimates of Ref. \cite{Kataev:1994rw}, which 
were in qualitative agreement with the estimate 
from Ref.\cite{Karshenboim:1993rt}.

\section{FEYNMAN CHALLENGE}

The problem of estimates 
of high order perturbative corrections to physical quantities 
was first formulated by R. Feynman. In the   talk at the 1961 Solvey Conference
he mentioned : ``As a special chellenge, is there any method of computing 
the anomalous moment of the electron which, on the first rough 
approximation, gives a fair approximation to the $\alpha$-term and a crude 
one to  $\alpha^2$, and when improved, increases the accuracy of the 
$\alpha^2$ term, yielding a rough estimate to $\alpha^3$ and beyond ?''
\cite{Feynman:1961}.
This question reveales  the useful features 
of  theoretical estimates. At the first stage  they may  give  the 
impression 
on the sign-structure of perturbative series, at the second stage 
are stimulating 
studies  of the effects, not included in these estimates, 
which when calculated 
and improved at the third stage are giving the final result for the whole 
correction.  

\subsection{Anomalous magnetic moment of muon: 8th order QED results}
The general QED expression for $a_{\mu}$ 
is :
\begin{eqnarray}
\nonumber 
a_{\mu}&=& a_e +A_2(m_{\mu}/m_e) \\
&+& A_2(m_{\mu}/m_{\tau}) + A_3(m_{\mu}/m_e,m_{\mu}/m_{\tau})
\end{eqnarray}
where 
\begin{equation}
\label{expr}
A_i=A_i^{(2)}\bigg(\frac{\alpha}{\pi}\bigg)+A_i^{(4)}
\bigg(\frac{\alpha}{\pi}\bigg)^2+A_i^{(6)}\bigg(\frac{\alpha}{\pi}\bigg)^3
+\dots
\end{equation}
and $i=1,2,3$. 
The first three  corrections to $a_e$ are known in the analytical 
form from the calculations of Refs. \cite{Schwinger:1948iu}-
\cite{Laporta:1996mq}. The updated 
value of the  8th order correction to $a_e$ was  
presented in Ref. \cite{Kinoshita:2005ti}.

The  dominant numerical values of  the terms $A_2^{(4)}$ and
$A_2^{(6)}$  are known  
and read \cite{Passera:2004bj,Kinoshita:2005ti} :
\begin{eqnarray}
A_2^{(4)}(m_{\mu}/m_e)&=&1.0942582887(104)~, \\ 
A_2^{(6)}(m_{\mu}/m_{e})&=& 22.86837936(22)~ .
\end{eqnarray}
Other terms in Eq. (\ref{expr}) are rather small 
and are of order $10^{-4}$- $10^{-5}$ \cite{Passera:2004bj,Kinoshita:2005ti}.
 The re-evaluation of the 8th order contributions to 
$a_{\mu}$ 
gives  the improved number \cite{Kinoshita:2004wi}, namely :
\begin{equation}
A_2^{(8)}(m_{\mu}/m_{e})= 132.6823(72) \\ 
\end{equation}
Notice, that the coefficients of $A_2(m_{\mu/m_e})$ are positive 
and their values are increasing. This happens due to the contributions 
of the powers of the relatively large  renormalization-group (RG) 
controllable terms with  ${\rm ln}(m_{\mu}/m_e)\approx 5.6$.
Moreover, beginning  from the  6th order the light-by-light-type 
diagrams with internal fermion loop are starting to manifest themselves 
\cite{Aldins:1970id}. 
Their typical contribution  are proportional to 
$\pi^2\rm{ln}(m_{\mu}/m_e)$--factors, which have non-RG origin and
are dominating in the expressions for the corresponding coefficients
of the 8th order correction. 
Thus, one may expect, that they will continue to dominate 
in higher orders also.

\subsection{10th order QED corrections 
to $a_{\mu}$}
The first estimate of the 10th-order correction to $a_{\mu}$ was given 
in Ref. \cite{Kinoshita:1990wp} on the basis of  rather preliminary 
numerical evaluation of the 10th-order diagrams with electron light-by-light 
subgraph and two one-loop electron vacuum polarization insertions into 
internal virtual photons, coupled to the  muon line. 
This estimate reads  \cite{Kinoshita:1990wp} 
\begin{equation}
\Delta_1(A_2^{(10)})\approx 570(140)~~~. 
\label{est1}
\end{equation}
However, there are 
at least 
two other sets of diagrams 
which were not taken into account in the estimate of Eq.(\ref{est1})
and may give sizable  contribution.
Among them is the light-by-light- type diagram, 
where one of three photons contains 
two-loop electron vacuum polarization insertion.  
Its  contribution was estimated in Ref. 
\cite{Karshenboim:1993rt} and reads 
\begin{equation}
\Delta_2(A_2^{(10)})\approx 176(35)~~
~.
\label{est2} 
\end{equation}
In the same work 
the contribution to $A_2^({10})$ 
of the diagram with electron loop, coupled to muon line by five 
photons, was estimates as \cite{Karshenboim:1993rt} :
\begin{equation}
\Delta_3(A_2^{(10)})\approx 185(85)~~. 
\label{est3}
\end{equation}
Eq.(\ref{est3}) includes 
theoretical and numerical information, gained
from Refs. \cite{Milstein}. 
Summing up the estimates of Eq. (\ref{est1}) - Eq.(\ref{est3}) one can get 
\cite{Karshenboim:1993rt}
\begin{equation}
\Delta_4(A^{(10)})= \Delta_1+\Delta_2 +
\Delta_3
~~\approx 930(170)~~.
\label{10est1}
\end{equation}

Another, more theoretical estimate, was made in Ref. \cite{Kataev:1994rw}.
It is  based on application of the  
scheme-invariant methods, namely the   
principle of minimal sensitivity 
\cite{Stevenson:1981vj} or  the effective charges method 
\cite{Grunberg:1982fw}. In the  estimates 
of Ref. \cite{Kataev:1994rw} the  
information on the values of lower-order contributions  
to $a_{\mu}$ (up to 8th order) and on the four-loop expression 
for  the QED $\beta$-function in the on-shell scheme \cite{Broadhurst:1992za} 
were used. The developed approach,
when  applied separately to the sets of
 non-light-by-light terms and the sum of light-by-light-type 
contributions,  gave  the following numbers \cite{Kataev:1994rw}
\begin{eqnarray}
\Delta_1^{ECH}(A_2^{(10)})&\approx&  50 \\ 
\Delta_2^{ECH}(A_2^{(10)})&\approx&  521~~~.  
\label{lbl}
\end{eqnarray}
Note, that Eq.(\ref{lbl}) contains  
the estimates for the sum of several
10th order contributions, 
including the ones, estimated separately  within other approaches 
in Eq.(\ref{est1}) and Eq. (\ref{est2}).  However, to obtain 
the final estimate  within this scheme-invariant method  it is also 
necessary to add the contribution of Eq. (\ref{est3}).
Thus the estimate of the 10th order QED correction to $a_{\mu}$, 
obtained in Ref.  \cite{Kataev:1994rw}, was 
\begin{equation}
\Delta_3^{ECH}(A^{(10)})= \Delta_1^{ECH}+ \Delta_2^{ECH}+
\Delta_3  \approx 750~.
\label{10est2}
\end{equation}
Within existing theoretical uncertainties the number of Eq. (\ref{10est2})
do not contradict to the one of Eq. (\ref{10est1}).  

However, quite recently more detailed 10th order results, 
based on the calculations of Kinoshita and Nio \cite{KN}, were
announced  \cite{Kinoshita:2005ti}. These results are:  
\begin{eqnarray}
\Delta_1(A_2^{(10)})&=& 629.1407(118) \\ 
\Delta_2(A_2^{(10)})&=& 181.1285(51) \\
\Delta_3(A_2^{(10)})&=& 86.69 
\label{new}
\end{eqnarray}
Kinoshita and Nio also 
calculated several other sets of  10th order diagrams 
diagrams, including the ones  
evaluated  previously   in 
Refs. \cite{Broadhurst:1992za}- \cite{Laporta:1994md}.  
The new estimate, which is based on the calculated  part
 of 9080 diagrams, contributing to the  
the 10th order 
QED contribution, is  \cite{Kinoshita:2005ti}:  
\begin{equation}
\Delta_{new}(A_2^{(10)})= 677(40)~~~.
\label{10est3}
\end{equation}
Note, that the 
calculations of the terms estimated in Eq. (\ref{est3})
are leading to the essential reduction of their
contribution into the 10th order correction 
to $a_{\mu}$ (compare  Eq. (\ref{new}) with 
Eq. (\ref{est3}).
Taking into account the effect 
of reduction of the contribution of 
$\Delta_3$ into Eq. (\ref{10est2})
we obtain a new estimate
 \begin{equation}
\Delta_{new}^{ECH}(A_2^{(10)}) \approx 658
\label{10est4}
\end{equation}
which is in perfect agreement with the  
estimate of Eq.(\ref{10est3}), based on explicit  
 calculations
of Ref. \cite{KN}.  We believe, that this good agreement is not the accident
and is demonstrating that both theoretical logic of scheme-invariant 
methods and the results of exact calculations are in good shape and 
are supporting each other. More detailed analysis of these results 
will be presented elsewhere. 

As to phenomenological 
consequence, the agreement of the preliminary partial results 
of  10th order calculations to $a_{\mu}$ with the scheme-invariant result 
of Eq. (\ref{10est3}) demonstrates, that the uncertainties 
of the 10th order QED contributions to $a_{\mu}$ are really small.
However, there is 
the  possibilities of decreasing current  theoretical uncertainties 
to  $a_{\mu}$. It can be done as the result 
of taking into account in the calculations of the hadronic 
vacuum polarization contributions  (for their evaluation see e.g. 
the  reviews of  
 Refs. \cite{Passera:2004bj},\cite{Jegerlehner:2003qp})
new data in the low energy region,  which  
will be  obtained soon at Novosibirsk $e^+e^-$ collider, and to rely 
on possible reconstruction of DAPHNE (Frascati) machine with the aim 
to measure 
the region in $e^+e^-$ -annihilation cross-section, complementary to the one, 
studied at  Novosibisrk and Bejing  colliders.

{\bf ACKNOWLEDGEMENTS}
This talk was prepared during the visit to ICTP (Trieste). 
I am grateful to the staff of this Center for 
providing excellent conditions for work. I also thank 
organizers of NuFact05 Workshop 
and V. Palladino in particular 
for invitation and hospitality in Frascati.
I am grateful to V.  Starshenko 
for the interest in the current status 
of different studies, related to $a_{\mu}$ and to T. Kinoshita for useful
correspondence.


\begin{thebibliography}{99}
\bibitem{Passera:2004bj}
  M.~Passera,
  J.\ Phys.\ G {\bf 31} (2005) R75.
\bibitem{Kinoshita:1990wp}
  T.~Kinoshita, B.~Nizic and Y.~Okamoto,
  Phys.\ Rev.\ D {\bf 41} (1990) 593.
\bibitem{Kinoshita:2002ns}
  T.~Kinoshita and M.~Nio,
  Phys.\ Rev.\ Lett.\  {\bf 90} (2003) 021803.
\bibitem{Kinoshita:2004wi}
  T.~Kinoshita and M.~Nio,
  Phys.\ Rev.\ D {\bf 70} (2004) 113001.
\bibitem{Milstein}
A.I.~Milstein and A.S. Ylkhovsky,
Phys. \ Lett.\ B {\bf 233}  (1989) 11
\bibitem{Karshenboim:1993rt}
  S.~G.~Karshenboim,
  Phys.\ Atom.\ Nucl.\  {\bf 56} (1993) 857
  [Yad.\ Fiz.\  {\bf 56N6} (1993) 252].
\bibitem{Kinoshita:2005ti}
 T.~Kinoshita,
  Nucl.\ Phys.\ Proc.\ Suppl.\  {\bf 144} (2005) 206.
\bibitem{Kataev:1994rw}
  A.~L.~Kataev and V.~V.~Starshenko,
  Phys.\ Rev.\ D {\bf 52} (1995) 402.
\bibitem{Feynman:1961}
R.~P. Feynman, in `` The Quantum Theory Fields'', Intescience Publishing 
Inc., NY,  1961. 
\bibitem{Schwinger:1948iu}
  J.~S.~Schwinger,
 Phys.\ Rev.\  {\bf 73} (1948) 416.
\bibitem{alpha2cor}
C.~Sommerfield, Phys. \ Rev. \ {\bf 107}  (1957) 328;\\
A.~ Petermann, Helv. \ Phys. \ Acta \ {\bf 3} (1957) 407;
M.~V.~Terentiev, J. \ Exp. \ Theor. \ Phys. \ {\bf 16} (1963) 444. 
\bibitem{Laporta:1996mq}
S.~Laporta and E.~Remiddi,
Phys.\ Lett.\ B {\bf 379} (1996) 283.
\bibitem{Kin67}
T. Kinoshita, Nuov.  Cim. \ B {\bf 51} (1967) 140.
\bibitem{Aldins:1970id}
J.~Aldins, S.~J.~Brodsky,  A.~J.~Dufner and T. Kinoshita,
Phys.\ Rev.\ D {\bf 1} (1970) 2378.
\bibitem{Stevenson:1981vj}
  P.~M.~Stevenson,
  Phys.\ Rev.\ D {\bf 23} (1981) 2916.
\bibitem{Grunberg:1982fw}
  G.~Grunberg,
  Phys.\ Rev.\ D {\bf 29} (1984) 2315.
\bibitem{KN}
T. Kinoshita and M. Nio, work in  progress.
\bibitem{Broadhurst:1992za}
  D.~J.~Broadhurst, A.~L.~Kataev and O.~V.~Tarasov,
  Phys.\ Lett.\ B {\bf 298} (1993) 445.
\bibitem{Kataev:1991az}
  A.~L.~Kataev,
  JETP Lett.\  {\bf 54} (1991) 602
  [Pisma Zh.\ Eksp.\ Teor.\ Fiz.\  {\bf 54} (1991) 600].
\bibitem{Kataev:1991cp}
  A.~L.~Kataev,
   Phys.\ Lett.\ B {\bf 284} (1992) 401.
\bibitem{Laporta:1994md}
  S.~Laporta,
  Phys.\ Lett.\ B {\bf 328} (1994) 522.
\bibitem{Jegerlehner:2003qp}
  F.~Jegerlehner,
  Nucl.\ Phys.\ Proc.\ Suppl.\  {\bf 126} (2004) 325.
\end{thebibliography}
\end{document}